# "Om" ॐ-Theory of Macroscopic Electromagnetism: Greener Vibes for Isotropy-Broken Media

Maxim Durach (mdurach@georgiasouthern.edu)

Center for Advanced Materials Science, Department of Biochemistry, Chemistry & Physics, Georgia Southern University, Statesboro, GA 30460, USA

**Abstract**

The applicability ranges of macroscopic and microscopic electromagnetisms are opposite. While microscopic electromagnetism deals with point sources, singular fields, and discrete atomistic materials, macroscopic electromagnetism concerns smooth average distributions of sources, fields, and homogenized effective metamaterials. Green's function method (GFM) involves finding fields of point sources and applying superposition principle to find fields of distributed sources. When utilized to solve microscopic problems GFM is perfectly within the applicability range. Extension of GFM to simple macroscopic problems is convenient, but not fully logically sound, since point sources and singular fields are technically not a subject of macroscopic electromagnetism. This explains the difficulty of both finding the Green's functions and applying superposition principle in complex isotropy-broken media, which are very different from microscopic environments. In this manuscript, we lay out a path to solution of macroscopic Maxwell's equations for distributed sources bypassing GFM, by introducing inverse approach and a method based on "Om" ॐ-potential which we describe here. To the researchers of electromagnetism this provides access to powerful analytical tools and a broad new space of solutions for Maxwell's equations.

1. **Introduction**

The main problem of electromagnetism is to be able to predict interaction between arbitrary charge distributions placed into arbitrary environments [1]. The path to solving this problem is most typically understood as finding fields of point sources. The fields of complex sources can be then obtained via superposition principle. This approach, known as Green's function method, is not only mathematically natural for singular microscopic fields but is grounded in physics of elementary particles which do not have dimensions according to relativity considerations [2]. The ubiquitous position of Green's functions in microscopic electromagnetism is best expressed in the essay by Julian Schwinger titled "The Greening of Quantum Field Theory: George and I" [3]. The "Greening" of macroscopic electromagnetism is less obvious, since photonics researchers do not deal with elementary particles or singular microscopic fields and do not claim the applicability of macroscopic photonics to elementary particles. Not unexpectedly, extension of the microscopic Green's function approach to macroscopic electromagnetism faces difficulties, due to fundamental differences between the corresponding sets of Maxwell's equations.

A lot of effort is invested into extending Green's function method to various macroscopic electromagnetic media; this cannot be considered very successful, however, at the rugged frontier of isotropy-broken media, due to the complexity and inherent non-locality of these media [4].

Although integral representation of Green's function can be obtained for the most general case of isotropy-broken media [5], closed-form expressions of dyadic Green's functions are only available for a limited set of relatively simple isotropy-broken media [6,7]. Another complication arises from utilizing superposition principle in the Green's function method, since finding fields created by non-point sources involves untenable integration even in the depolarization dyadics approximation [7]. Finding fields of non-point sources in isotropic media in many cases relies on symmetries of those sources [1]. This can be extended to isotropy-broken media by means of spectral eigenfunction representations, which, however, results in infinite series, requiring truncation [7,8].

In this manuscript we introduce two approaches to directly obtain fields created by a very broad class of sources immersed in generic isotropy-broken media. First, we apply inverse approach to the inhomogeneous Helmholtz equation for the vector potential, to obtain sources that create desired vector potentials. In the second approach we seek inspiration from the teachings of Hinduism philosophy about the primordial fore-sound of the universe encompassing all Creation. We introduce the "Om" ॐ-potential that underlies both sources and fields and provides for the direct method of evaluation of the solutions of macroscopic Maxwell's equations in isotropy-broken media. Please note that introduction of auxiliary vector fields to aid solution or analysis of Maxwell equations is not unprecedented in history of science as exemplified by the scalar potential, vector potential, Hertz potential [9], and Beltrami fields [10]. The power of our methods is demonstrated by the mappings we uncover between different sources that create identical potential across all materials and between field-source pairs which come from the same "Om" ॐ in materials as they transition between symmetries, topology classes, and so on.

## 2. Helmholtz Equation and Green's Functions in Isotropy-Broken Media

Macroscopic fields satisfy Maxwell's equations

$$\nabla \times \boldsymbol{H} - \frac{1}{c}\frac{\partial \boldsymbol{D}}{\partial t} = \frac{4\pi}{c}\boldsymbol{j}, \quad \nabla \times \boldsymbol{E} + \frac{1}{c}\frac{\partial \boldsymbol{B}}{\partial t} = 0 \quad (1)$$

The constitutive relations are generally expressed as

$$\begin{pmatrix}\boldsymbol{D}\\\boldsymbol{B}\end{pmatrix} = \widehat{M}\begin{pmatrix}\boldsymbol{E}\\\boldsymbol{H}\end{pmatrix} = \begin{pmatrix}\hat{\epsilon} & \hat{X}\\\hat{Y} & \hat{\mu}\end{pmatrix}\begin{pmatrix}\boldsymbol{E}\\\boldsymbol{H}\end{pmatrix} \text{ or } \begin{pmatrix}\boldsymbol{D}\\\boldsymbol{H}\end{pmatrix} = \begin{pmatrix}\hat{C}_{DE} & \hat{C}_{DB}\\\hat{C}_{HE} & \hat{C}_{HB}\end{pmatrix}\begin{pmatrix}\boldsymbol{E}\\\boldsymbol{B}\end{pmatrix} = \begin{pmatrix}(\hat{\epsilon} - \hat{X}\hat{\mu}^{-1}\hat{Y}) & \hat{X}\hat{\mu}^{-1}\\-\hat{\mu}^{-1}\hat{Y} & \hat{\mu}^{-1}\end{pmatrix}\begin{pmatrix}\boldsymbol{E}\\\boldsymbol{B}\end{pmatrix} \quad (2)$$

Under Weyl gauge the relationship between fields and the vector potential reduce to $\boldsymbol{B} = \nabla \times \boldsymbol{A}$ and $\boldsymbol{E} = -\frac{1}{c}\frac{\partial \boldsymbol{A}}{\partial t}$. Combining Maxwell's equations and constitutive relations, we obtain the wave equation for the vector potential in isotropy-broken media

$$\left(\frac{1}{c}\frac{\partial}{\partial t}, -\nabla \times \hat{I}\right)\begin{pmatrix}\hat{C}_{DE} & \hat{C}_{DB}\\\hat{C}_{HE} & \hat{C}_{HB}\end{pmatrix}\begin{pmatrix}-\frac{1}{c}\frac{\partial}{\partial t}\\\nabla \times \hat{I}\end{pmatrix}\boldsymbol{A} = \hat{L}\left(\nabla, \frac{1}{c}\frac{\partial}{\partial t}\right)\boldsymbol{A} = -\frac{4\pi}{c}\boldsymbol{j}_s \quad (3a)$$

or

$$-\hat{C}_{DE}\frac{1}{c^2}\frac{\partial^2 A}{\partial t^2} + \hat{C}_{DB}\nabla \times \frac{1}{c}\frac{\partial A}{\partial t} + \nabla \times \hat{C}_{HE}\frac{1}{c}\frac{\partial A}{\partial t} - \nabla \times \hat{C}_{HB}\nabla \times A = -\frac{4\pi}{c}j_s \quad (3b)$$

More conventionally it is expressed as

$$-\left(\nabla \times \hat{I} - \frac{1}{c}\frac{\partial}{\partial t}\hat{X}\right)\hat{\mu}^{-1}\left(\nabla \times \hat{I} + \hat{Y}\frac{1}{c}\frac{\partial}{\partial t}\right)A - \hat{\epsilon}\frac{1}{c^2}\frac{\partial^2 A}{\partial t^2} = -\frac{4\pi}{c}j_s \quad (3c)$$

Transforming into the Fourier domain as $\hat{L}\left(\nabla, \frac{1}{c}\frac{\partial}{\partial t}\right) \to \hat{L}(i\mathbf{k}, -ik_0)$ results in the Helmholtz operator for isotropy-broken media

$$\hat{L}(i\mathbf{k}, -ik_0) = (k_0, \mathbf{k} \times \hat{I})\begin{pmatrix}\hat{C}_{DE} & \hat{C}_{DB} \\ \hat{C}_{HE} & \hat{C}_{HB}\end{pmatrix}\begin{pmatrix}k_0 \\ \mathbf{k} \times \hat{I}\end{pmatrix} = (\mathbf{k} \times \hat{I} + k_0\hat{X})\hat{\mu}^{-1}(\mathbf{k} \times \hat{I} - k_0\hat{Y}) + k_0^2\hat{\epsilon} \quad (4)$$

The vector potential can be now expressed as $\mathbf{A}(\mathbf{k}, k_0) = -\frac{4\pi}{c}\frac{\text{adj}\,\hat{L}}{|\hat{L}|}\mathbf{j}(\mathbf{k}, k_0)$ and

$$\mathbf{A}(\mathbf{r}, k_0) = -\frac{4\pi}{c}\int \frac{d^3k}{(2\pi)^3}\frac{\text{adj}\,\hat{L}}{|\hat{L}|}\mathbf{j}(\mathbf{k}, k_0)\exp(i\mathbf{k}\mathbf{r}), \quad (5)$$

where the $|\hat{L}| = \frac{1}{k_0^2}\sum_{i+j+l+m=4}[\alpha_{ijlm}k_x^i k_y^j k_z^l k_0^m]$ is the determinant of the operator Eq. (4), with $\alpha_{ijlm}$ being the Tamm-Rubilar tensor [11,12], and the corresponding adjoint operator is $\text{adj}\,\hat{L} = \frac{1}{k_0^4}\sum_{i+j+l+m=4}[\hat{\beta}_{ijlm}k_x^i k_y^j k_z^l k_0^m]$, where $\hat{\beta}_{ijlm}$ are 3x3 matrices.

The usual approach is to use the *superposition principle* to express Eq. (5) in terms of dyadic Green's function $\hat{G}(\mathbf{r}, \mathbf{r}')$

$$\mathbf{A}(\mathbf{r}) = -\frac{4\pi}{c}\int d\mathbf{r}'\,\hat{G}(\mathbf{r}, \mathbf{r}')\,\mathbf{j}(\mathbf{r}') \quad (6)$$

Correspondingly, from Eqs. (5)-(6), the dyadic Green's function can be expressed as [5,7]

$$\hat{G}(\mathbf{r}, \mathbf{r}') = \int \frac{d^3k}{(2\pi)^3}\frac{\text{adj}\,\hat{L}}{|\hat{L}|}\exp(i\mathbf{k}(\mathbf{r}-\mathbf{r}')), \quad (7)$$

The problem of finding fields created by arbitrary sources is reduced this way to finding the Green's function, which itself is a response to *a delta-functional point source*, and always has a singular part [7]

$$\hat{L}(\nabla, -ik_0)\,\hat{G}(\mathbf{r}, \mathbf{r}') = \hat{I}\,\delta(\mathbf{r} - \mathbf{r}') \quad (8)$$

As described in the introduction, in general, both finding the Green's function Eq. (7)-(8) and utilizing the superposition principle, Eq. (6), is a challenge in macroscopic electromagnetism and is unnatural, due to the limited validity of point sources and singular fields in macroscopic environments.

## 3. The "Om" ॐ potential

To bypass the complications related to GFM we introduce differential operators based on Fourier space operators $|\hat{L}|$ and adj $\hat{L}$

$$D(\partial_x, \partial_y, \partial_z) = \frac{1}{k_0^2}\sum_{i+j+l+m=4}(-i)^{i+j+l}[\alpha_{ijlm}\partial_x^i\partial_y^j\partial_z^l k_0^m] \tag{9a}$$

$$\hat{U}(\partial_x, \partial_y, \partial_z) = \frac{1}{k_0^4}\sum_{i+j+l+m=4}(-i)^{i+j+l}[\hat{\beta}_{ijlm}\partial_x^i\partial_y^j\partial_z^l k_0^m] \tag{9b}$$

Note that differential operators $D$ and $\hat{U}$ have constant coefficients in homogeneous media and, therefore, commute. We recast Eq. (5) as

$$D(\partial_x, \partial_y, \partial_z)A(r) = -\frac{4\pi}{c}\hat{U}(\partial_x, \partial_y, \partial_z)j(r) \tag{10}$$

Instead of representing the source $j(r)$ as a superposition of point charges as is done in the Green's function method, we express the source via the underlying "Om" ॐ potential vector field

$$j(r) = D(\partial_x, \partial_y, \partial_z)\,ॐ(r)\,, \tag{11}$$

From Eq. (10) an expression for the vector potential corresponding to source current Eq. (11) can be obtained as

$$A(r) = -\frac{4\pi}{c}\hat{U}(\partial_x, \partial_y, \partial_z)\,ॐ(r) \tag{12}$$

where Devanagari script "Om" ॐ$(r)$ is a vector field which underlies both the source $j$ in Eq. (11) and the vector potential $A$ in Eq. (12) in a unified paradigm of Eqs. (10)-(12).

Note that for arbitrary source the underlying "Om" vector field ॐ$(r)$ can be found as

$$ॐ(r) = \int dr'\, g_ॐ(r,r')\,j(r'),$$

where the scalar "Om" Green's function is $g_ॐ(r,r')$

$$g_ॐ(r,r') = \int \frac{d^3k}{(2\pi)^3}\frac{1}{|\hat{L}|}\exp(ik(r-r'))$$

The summary of relationships between the "Om" ॐ potential, vector potential $A$, and sources $j$ is shown in Fig. 1.

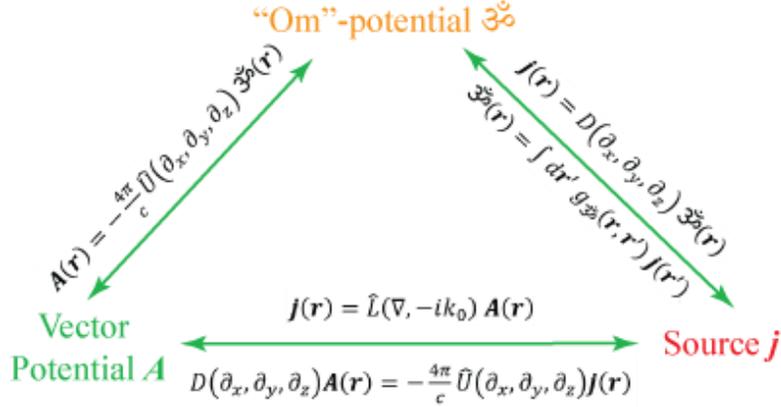

Fig. 1. The schematic of the relations between the sources $j(r)$, vector potentials $A(r)$, and the "Om" ॐ-potential introduced in this manuscript.

In vacuum the operator $\hat{L}$ has the following properties

$$\hat{L}(i\boldsymbol{k},-ik_0) = (k_0, \boldsymbol{k}\times\hat{I})\begin{pmatrix} k_0 \\ \boldsymbol{k}\times\hat{I} \end{pmatrix} = (\boldsymbol{k}\times\hat{I})(\boldsymbol{k}\times\hat{I}) + k_0^2\hat{I} = \boldsymbol{kk} + (k_0^2 - k^2)\hat{I} \quad (6)$$

$$|\hat{L}| = k_0^2(k^2-k_0^2)^2, \quad \text{adj}\,\hat{L} = (k^2-k_0^2)(\boldsymbol{kk}+k_0^2\hat{I})$$

This means that for vacuum Eq. (10) can be rewritten as

$$(\nabla^2 + k_0^2)^2 \boldsymbol{A}(\boldsymbol{r}) = \frac{4\pi}{c}(\nabla^2+k_0^2)\left(\hat{I} - \frac{1}{k_0^2}\nabla\nabla\right)\boldsymbol{j}(\boldsymbol{r})$$

$$D_{vac}(\partial_x,\partial_y,\partial_z) = k_0^2(\nabla^2+k_0^2)^2, \quad \hat{U}_{vac}(\partial_x,\partial_y,\partial_z) = (\nabla^2+k_0^2)(k_0^2\hat{I} - \nabla\nabla)$$

For a point source at $\boldsymbol{r}_0$ polarized in direction $\hat{\boldsymbol{e}}$ in vacuum $\boldsymbol{j} = \hat{\boldsymbol{e}}\,\delta(\boldsymbol{r}-\boldsymbol{r}_0)$ the "Om" vector field $\text{ॐ}(\boldsymbol{r})$ is a spherical wave propagating from the source location

$$\text{ॐ}_{vac-point}(\boldsymbol{r}) = (\nabla^2+k_0^2)\,\hat{\boldsymbol{e}}\,\frac{e^{ik_0|\boldsymbol{r}-\boldsymbol{r}_0|}}{4\pi|\boldsymbol{r}-\boldsymbol{r}_0|} = \hat{\boldsymbol{e}}\,e^{ik_0|\boldsymbol{r}-\boldsymbol{r}_0|}$$

### 4. Inverse Helmholtz equation Method

The first method to find solutions of Eq. (5) relies on inverse approach to the Helmholtz equation

$$\boldsymbol{j}(\boldsymbol{r}) = \hat{L}(\nabla,-ik_0)\,\boldsymbol{A}(\boldsymbol{r}), \qquad (13)$$

where instead of looking for vector potential $A(r)$ for a given source $j(r)$, we set the vector potential $A(r)$ and obtain sources $j(r)$, which create the desired vector potential.

To proceed, we utilize the Hermite functions $\phi_{n_x n_y n_z}(\mathbf{r}) = \psi_{n_x}(x)\psi_{n_y}(y)\psi_{n_z}(z)$, which are eigenfunctions of the quantum harmonic oscillator $\psi_{n_x}(x) = \left(2^{n_x} n_x! \sqrt{\pi} w_x\right)^{-\frac{1}{2}} e^{-\frac{x^2}{2w_x^2}} H_n(x/w_x)$. The del operator applied to Hermite functions is

$$\nabla \psi_{n_x n_y n_z}(\mathbf{r}) = \frac{1}{\sqrt{2}} \left\{ (\hat{a}_x - \hat{a}_x^+)/w_x, (\hat{a}_y - \hat{a}_y^+)/w_y, (\hat{a}_z - \hat{a}_z^+)/w_z \right\} \psi_{n_x n_y n_z}(\mathbf{r})$$

where the ladder operators $\hat{a}_x \psi_n = \sqrt{n+1}\psi_{n+1}$, $\hat{a}_x^+ \psi_n = \sqrt{n}\psi_{n-1}$.

For a vector potential in vacuum polarized in direction $\hat{x}$ and given by $\mathbf{A}(\mathbf{r}) = \hat{x}\,\psi_{000}(\mathbf{r})$ the source can be found as

$$\mathbf{j}(\mathbf{r}) = \hat{L}(\nabla, -ik_0)\mathbf{A}(\mathbf{r}) = \{(k_0^2 + \nabla^2)\hat{I} - \boldsymbol{\nabla}\boldsymbol{\nabla}\}\mathbf{A}(\mathbf{r}) =$$

$$= \frac{1}{\sqrt{2}} \frac{1}{w^2} \{\sqrt{2}(k_0^2 w^2 - 1)\psi_{000} + \psi_{020} + \psi_{002}, -\psi_{110}, -\psi_{101}\} \quad (14)$$

If the vector potential $\mathbf{A}(\mathbf{r})$ is fixed, the only material-dependent factor in the RHS of Eq. (13) is the operator $\hat{L}(\nabla, -ik_0)$. This allows us to create a cross-material mapping between sources $\mathbf{j}(\mathbf{r})$ which create the same vector potential in different media. To demonstrate this, we consider a material with $\widehat{M}_\kappa = (1-\kappa)\hat{1} + \kappa\widehat{M}$, where matrix $\widehat{M}$ is color-coded in Fig. 2(e). As $\kappa$ is changed from 0 to 1 the material passes through several topological transitions (see Refs. [13,14]) from non-hyperbolic, to mono-hyperbolic [Fig. 2(a)], to bi-hyperbolic [Fig. 2(b)], to tri-hyperbolic [Fig. 2(c)], to tetra-hyperbolic [Fig. 2(d)].

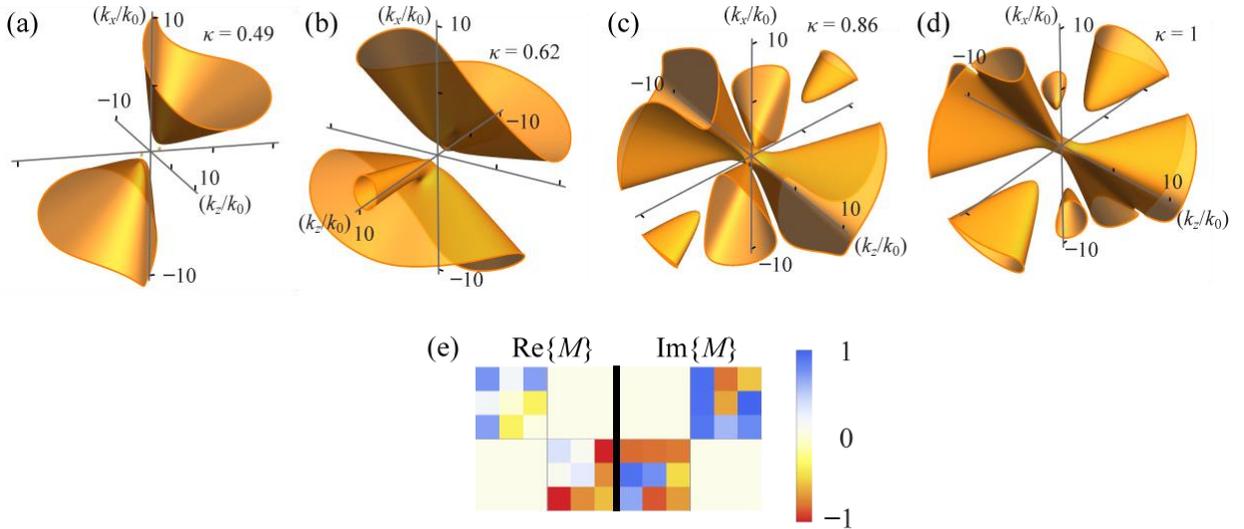

Fig. 2. Topological transitions of material $\widehat{M}_\kappa = (1-\kappa)\hat{1} + \kappa\widehat{M}$ for different $\kappa$.

In Fig. 3 we show how the source $j(r)$ required to create vector potential $A(r) = \hat{x}\,\psi_{000}(r)$ changes as $\kappa$ is changed. The leftmost panels of Fig. 3 correspond to vacuum $\kappa = 0$ and follow Eq. (14). One can see that the source distribution is deformed and rotated as the materials are passing through topological transitions shown in Fig. 2.

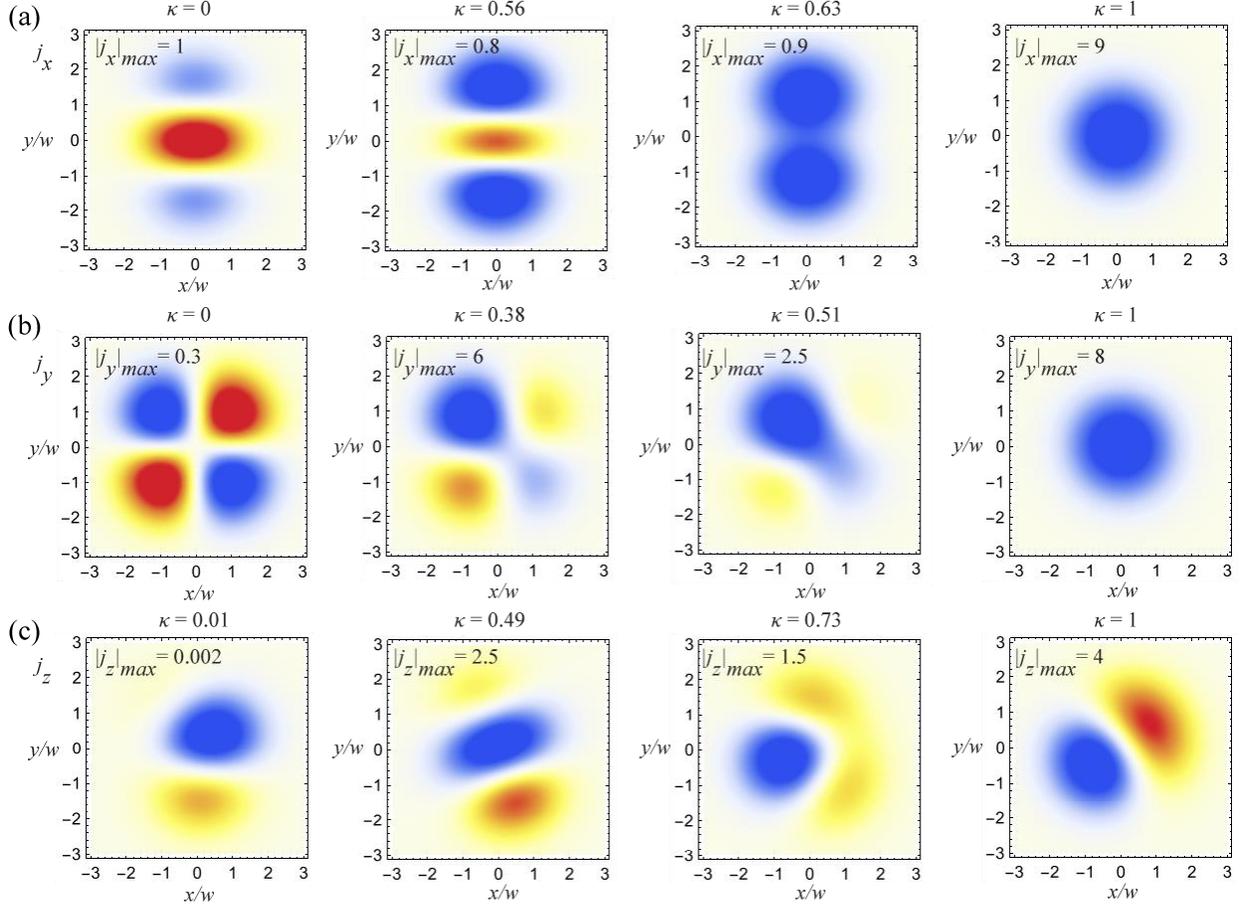

Fig. 3. The x-y plane cross-section of the sources $j(r)$ needed to create potential $A(r) = \hat{x}\,\psi_{000}(r)$ in different materials $\widehat{M}_\kappa$. Panel (a) shows x-component $j_x$; (b) $j_y$; (c) $j_z$.

### 1. The "Om" ॐ potential Method

The second method to find solutions of Eq. (5) is to use Eqs. (11)-(12). We select the "Om" ॐ$(r)$-potential and find the corresponding source $j(r)$ and vector potential $A(r)$. If the "Om" ॐ$(r)$-potential is fixed, the only material-dependent factors in the RHS of Eqs. (11)-(12) are operators $D(\partial_x, \partial_y, \partial_z)$ and $\widehat{U}(\partial_x, \partial_y, \partial_z)$. This creates cross-material mapping between the source-vector potential pairs $j(r)$ and $A(r)$, which correspond to the same "Om" ॐ$(r)$-potential as material is modified.

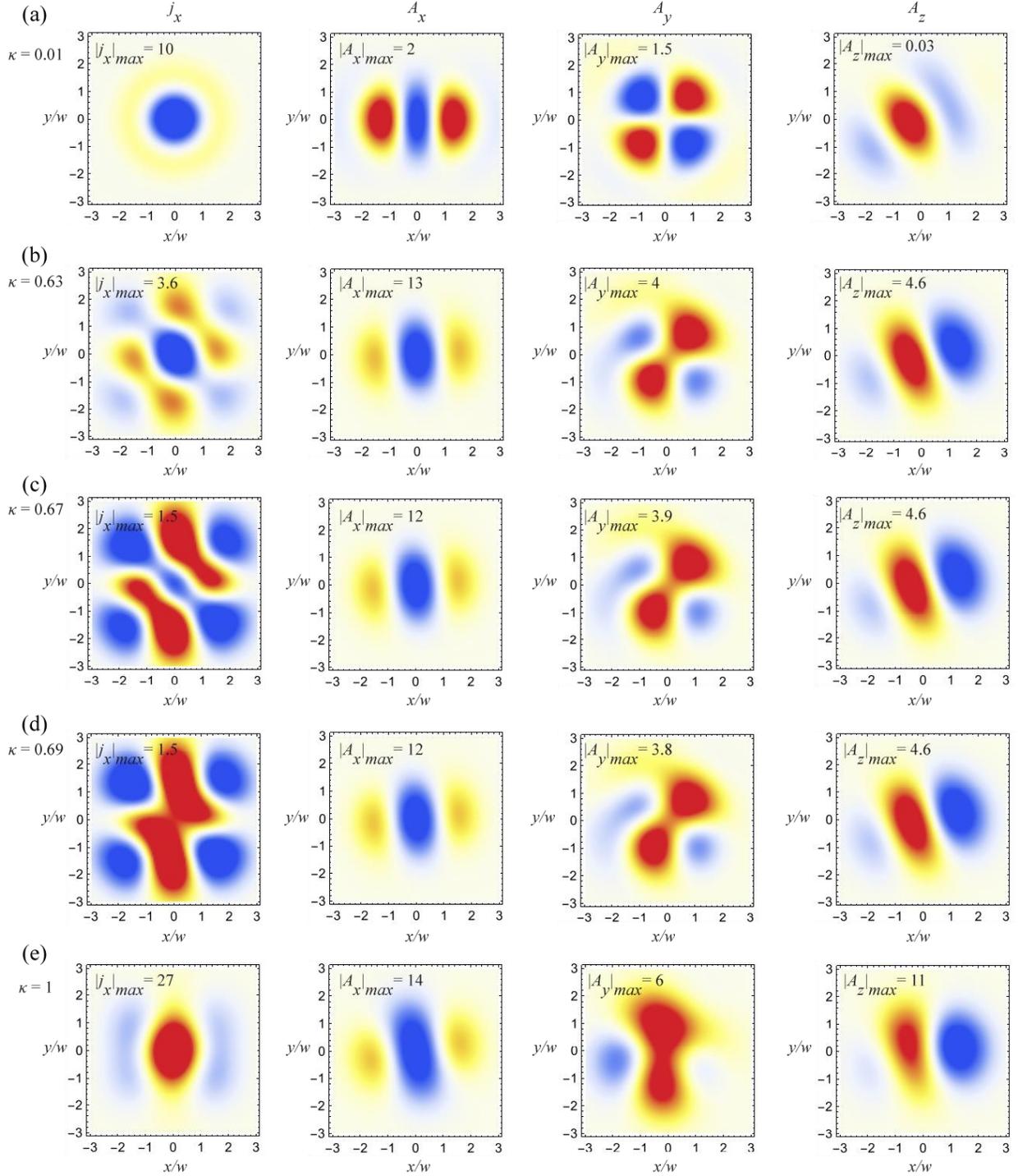

Fig. 4. The x-y plane cross-sections of the x-component of the sources $j_x$ (leftmost panels) and the components of the vector potential $\boldsymbol{A}(\boldsymbol{r})$ (three rightmost panels) for the "Om"-potential given by $\check{\mathcal{P}}(\boldsymbol{r}) = \hat{\boldsymbol{x}}\,\psi_{000}(\boldsymbol{r})$ for different materials $\widehat{M}_\kappa$. In panel (a) $\kappa = 0.01$; (b) $\kappa = 0.63$; (c) $\kappa = 0.67$; (d) $\kappa = 0.69$; (e) $\kappa = 1$.

In Fig. 4 we show the x-y cross-sectional distributions of the sources $\boldsymbol{j}(\boldsymbol{r})$ and vector potentials $\boldsymbol{A}(\boldsymbol{r})$ corresponding to $\breve{\text{ॐ}}(\boldsymbol{r}) = \hat{\boldsymbol{x}} \psi_{000}(\boldsymbol{r})$ for different materials $\widehat{M}_\kappa$. We see drastic modifications of both $\boldsymbol{j}(\boldsymbol{r})$ and $\boldsymbol{A}(\boldsymbol{r})$, which undergo both deformation and rotation. Interestingly, the rate of change in x-y cross-sections of $\boldsymbol{j}(\boldsymbol{r})$ and $\boldsymbol{A}(\boldsymbol{r})$ is not the same as $\kappa$ is changed. For $\kappa = 0.63 - 0.69$, when the material is in the topological transition into bi-hyperbolic phase, the source $\boldsymbol{j}(\boldsymbol{r})$ is strongly modified as can be seen from leftmost panels in Fig. 3 (b)-(d). At the same time the vector potential $\boldsymbol{A}(\boldsymbol{r})$ has minimal changes in the same range of $\kappa$.

In conclusion, the Green's function method with point sources and singular fields is natural to microscopic electromagnetism and fundamentally stems from the properties of dimensionless elementary particles. Extension of GFM to macroscopic electromagnetism faces obvious and fundamental challenges. In this manuscript we demonstrate that solutions to problems of macroscopic electromagnetism can be found without the use of Green's functions by introducing two new approaches: the inverse Helmholtz equation method and the "Om" ॐ potential method.